# Effect of nano-carbon particle doping on the flux pinning properties of MgB$_2$ superconductor

S. Soltanian[1], J. Horvat[1], X.L. Wang[1], P. Munroe[2] and S.X. Dou[1]

1. Institute for Superconducting and Electronic Materials, University of Wollongong, Northfields Ave. NSW 2522, Australia
2. Electron Microscope Unit, University of New South Wales, NSW 2052, Australia

**Abstract**
Polycrystalline MgB$_{2-x}$C$_x$ samples with x=0.05, 0.1, 0.2, 0.3, 0.4 nano-particle carbon powder were prepared using an in-situ reaction method under well controlled conditions to limit the extent of C substitution. The phases, lattice parameters, microstructures, superconductivity and flux pinning were characterized by XRD, TEM, and magnetic measurements. It was found that both the *a*-axis lattice parameter and the T$_c$ decreased monotonically with increasing doping level. For the sample doped with the highest nominal composition of x=0.4 the T$_c$ dropped only 2.7K. The nano-C-doped samples showed an improved field dependence of the J$_c$ compared with the undoped sample over a wide temperature range. The enhancement by C-doping is similar to that of Si-doping but not as strong as for nano-SiC doped MgB2. X-ray diffraction results indicate that C reacted with Mg to form nano-size Mg$_2$C$_3$ and MgB$_2$C$_2$ particles. Nano-particle inclusions and substitution, both observed by transmission electron microscopy, are proposed to be responsible for the enhancement of flux pinning in high fields.

**Introduction**

The effect of C-doping on superconductivity in MgB$_2$ compound has been studied by several groups [1-8]. The results on C solubility and the effect of C-doping on T$_c$ reported so far vary significantly from no substitution to 15% of B substituted by C, while the decrease in T$_c$ ranged up to 17K at the highest substitution level [9,10]. The significant differences among the studies on C-substitution are attributable to the fabrication techniques and precursor materials used. It appears that lower sintering temperatures (e.g. 700$^o$C) and short sintering times result in an incomplete reaction and hence lower C solubility in MgB$_2$. The mixing procedure applied to the precursor materials may also contribute to inhomogeneity in the final product. It is difficult to precisely determine the C solubility in the lattice, as the lattice parameters can also be affected by the change of stoichiometry in MgB$_2$, because excess C extracts Mg and B to form MgB$_2$C$_2$. Recently, Ribeiro et al. used Mg and B$_4$C as precursors to synthesize C doped MgB$_2$ by sintering at 1200$^o$C for 24 hours [9]. Their samples appeared to be homogeneous. A neutron diffraction study confirmed that the most likely solubility of C in MgB$_2$ is up to around 10% of B positions [10]. This gives the largest drop in both T$_c$ (=22K) and the *a*-axis lattice parameter.

The studies on C doping into MgB$_2$ have thus far only focused on the effect on superconductivity. From the applications point of view, the effect of C doping on the flux pinning properties is also important. In this paper, we report on the effects of C doping on the flux pinning and critical current density in MgB$_2$. It is clear from previous work that complete substitution causes a drastic depression in T$_c$, which is very undesirable for improving J$_c$ at high temperatures. In order to explore the potential applications of MgB$_2$ at around 20K or above it is essential to maintain the T$_c$ and, at the same time, to enhance the J$_c$ performance in magnetic fields. So, based on the previous work on the in-situ reaction and fast formation of MgB$_2$ [11,12] we designed synthesis conditions that limit the degree of C substitution but can introduce effective pinning centers into MgB$_2$.

**Experimental**

Polycrystalline samples of $MgB_{2-x}C_x$ were prepared through a reaction in-situ process [11,12]. High purity powders of magnesium (99%), amorphous boron (99%) and carbon nano-particles (with a particle size of about 20nm) were weighed out according to the nominal atomic ratio of $MgB_{2-x}C_x$ with x = 0, 0.05, 0.1, 0.2, 0.3, 0.4 and well-mixed through grinding. The powders were pressed into pellets of 10 mm in diameter and 3 mm in thickness using a hydraulic press. The pellets were sealed in Fe tubes, then heat treated at 770 $^o$C for 30min in flowing high purity Ar. This was followed by a furnace cooling to room temperature. An un-doped sample was also made under the same conditions for use as a reference sample. The phase and crystal structure of all the samples was obtained from X-ray diffraction (XRD) patterns using a Philips (PW1730) diffractometer with Cu$K\alpha$ radiation. Si powder was used as a standard reference to calculate the lattice parameters. The grain morphology and microstructure were also examined by scanning electron microscope (SEM) and transmission electron microscope (TEM).

The magnetization was measured over a temperature range of 5 to 30 K using a Physical Properties Measurement System (PPMS, Quantum Design) in a time-varying magnetic field of sweep rate 50 Oe/s and amplitude 9T. Bar shaped samples with a size of about 4 x 3 x 0.5 mm$^3$ were cut from each pellet for magnetic measurements. The magnetic measurements were performed by applying the magnetic field parallel to the longest sample axis. The magnetic $J_c$ was calculated from the height $\Delta M$ of the magnetization loop (M-H) using the Bean model where $J_c$=20 $\Delta M/[a/(1-a/3b)]$, where a and b are the dimensions of the sample perpendicular to the direction of applied magnetic field with a<b. $J_c$ versus magnetic field has been measured up to 8.5 T for the samples at 5 K, 10 K, 20 K, and 30 K. The low field $J_c$ below 10 K cannot be measured due to flux jumping. The $T_c$ was determined by measuring the real part of the ac susceptibility at a frequency of 117 Hz and an external magnetic field of 0.1 Oe. $T_c$ was defined as the onset of the diamagnetism.

**Results and discussion:**

Fig. 1 shows the XRD patterns of $MgB_{2-x}C_x$ samples for x=0, 0.05, 0.1, 0.2, 0.3 and 0.4 as well as the XRD pattern of the starting C powder. An Si standard was used for all runs. It can be seen that there are no diffraction peaks for C powder, indicating that this powder is amorphous. Thus, there is no peak related to C in the XRD patterns of the C-doped samples, and the amount of the un-reacted C powder is not clear. The undoped samples consist of a main phase, $MgB_2$, with minor phases of MgO (<5%) and $MgB_4$. In the C-doped samples extra peaks appear as impurity phases. These peaks can be indexed as $Mg_2C_3$ and $MgB_2C_2$, which increase as the doping level increases.

More accurate XRD examinations were performed to evaluate the lattice parameters. The XRD patterns are shown in Fig. 2. Note that the position of the (100) peak shifts continuously to higher angles with increasing C doping level, indicating a decrease in the a-axis lattice parameter. However the position of the (002) peak remains unchanged with increasing C-doping level, indicating that C-doping does not affect the c axis. The changes in crystal lattice parameters deduced from the x-ray diffraction patterns of the samples as well as the lattice parameters extracted from the previously published studies by Maurin et al. [6] and Avdeev et al. [10] are shown in Fig. 3. As can be seen, the in-plane (*a* axis) lattice parameter decreases monotonically as the C doping level decreases from 3.087A$^o$ to 3.076A$^o$ at x=0 and x=0.4 respectively. This can be understood because the average size of the C ion (0.077nm) is smaller than the B (0.082nm). However, we are unable to see any significant change in the inter layer (*c*-axis) lattice parameter. This is in agreement with recent work, indicating that carbon is substituted in the boron honeycomb layer and does not change the interlayer distance in the $MgB_2$ crystal. However, the change in the *a* lattice parameter even for x=0.4 is considerably less than the *a* axis contraction from 3.085 to 3.052 due to 10% carbon doping [10]. This indicates that the carbon powder in our samples is only partially substituted in the B position due to the low sintering temperature and short sintering time. The C mostly reacted with Mg and B to form $Mg_2C_3$ and $MgB_2C_2$ or remained in an un-reacted form.

Fig. 4 shows the transition temperature ($T_c$) for the doped and undoped samples determined by ac susceptibility measurements. The $T_c$ onset for the undoped sample (~ 38.5 K) is almost the same as that reported by a number of groups. For the doped samples, the $T_c$ decreases with increasing doping level. Despite the large amount of non-superconducting phases present, the $T_c$ only drops slightly, 2.7K at a high C doping level of x=0.4 ( which represents 20% C doping). This result is in contrast to the previously reported results in which the $T_c$ was depressed about 17K in the 10% C substituted sample, as we can see in the inset of Fig. [10]. Once again, these results suggest that only a small amount of C powder was substituted in the B position in our samples, consistent with the lattice contraction. Because the C doping has little effect on the $T_c$, the partial substitution and partial addition of nano-carbon particles may enhance flux pinning within a wide range of temperatures.

Fig. 5 (a-d) shows the $J_c(H)$ curves for $MgB_2$ doped and undoped samples at 5 K, 10K, 20K and 30 K. It should be noted that at 5K, 10K and 20K, all the $J_c(H)$ curves for doped samples show a crossover with the undoped sample at higher fields except for the sample doped with x=0.4 at 20K. Because the C doping reduces $T_c$, only the $J_c(H)$ curve for the C-doped sample with x=0.05 shows the crossover with the undoped sample at 30K. Fig. 6 shows the irreversibility field, $H_{irr}$ versus temperature for all the doped and undoped samples. Here, we defined $H_{irr}$ as the field where $J_c$ drops to 100 A/cm$^2$. The improvement in $H_{irr}$ for all the C doped samples is consistent with the $J_c(H)$.

In order to understand the mechanisms for the enhancement of flux pinning in the nano-C doped samples a TEM study was performed. Fig. 7 shows a typical TEM image for the C-doped sample at x=0.05 (Fig. 7(a)) and x=0.1 (Fig.7 (b)). Note that the $MgB_2$ grains are approximately 100 -200nm long and 50-100nm wide.  It is evident that although the doping levels of x=0.05 and x=0.1 are well below the C solubility limit, there are noticeable amounts of precipitates which may be unreacted C and the impurity phases $Mg_2C_3$ and $MgB_2C_2$. These precipitates are uniformly distributed within the matrix and have a diameter of 5nm to 10nm. Many are included in the grains as fine inclusions. The density and amount of these inclusions increase with increasing doping level. The size of these inclusions matches very well the coherence length of $MgB_2$. Thus, it is believed that the high density of nano-inclusions is responsible for the enhanced flux pinning in the C-doped samples

In previous papers, we reported the results on nano-SiC and nano-Si doping into $MgB_2$ [13,14]. Compared to the un-doped sample, $J_c$ for the 10wt% SiC-doped sample increased by more than an order of magnitude in magnetic fields. Nano-Si particles showed a pinning enhancement, but it was not as strong as with SiC. Fig. 8 compares the normalized $J_c(H)$ and $H_{irr}$ for nano-SiC [13], nano-Si [14] and nano-C doped $MgB_2$ at 20K.  Note that C and Si doping gave the same level of enhancement over the undoped sample, while SiC-doped $MgB_2$ remained the best of all the forms of $MgB_2$ reported thus far. The extent of enhancement of $J_c$ and flux pinning due to the nano-SiC doping has exceeded that achieved by proton irradiation [15] and oxygen alloying [16]. The mechanism for the strong enhancement in nano-SiC doped $MgB_2$ remains an open question, as there are similar types of nano-inclusions in the SiC, C and Si doped samples. It is evident that inclusions or additives can not explain the differences and some other pinning mechanism such as structure modulation induced by substitution may be responsible for the extra pinning enhancement in SiC-doped $MgB_2$.

**Conclusions**

The effect of C doping on lattice parameters, $T_c$, $J_c$ and flux pinning in $MgB_2$ was investigated under the conditions of limited C substitution for B. It was found that both the *a*-axis lattice parameter and the $T_c$ decreased monotonically with increasing doping level. For the sample doped with the highest nominal composition of x=0.4 the $T_c$ dropped only 2.7K. The nano-C-doped samples showed an improved field dependence of the $J_c$ over a wide temperature range compared with the undoped sample. X-ray diffraction and TEM studies indicate that C reacted with Mg to form $Mg_2C_3$ and $MgB_2C_2$ with nano-dimensions. Nano-particle inclusions and substitution, both observed by transmission electron microscopy, are proposed to be responsible for the enhancement of flux pinning in high fields.

Acknowledgment

The authors thank Dr. T. Silver for her helpful discussion. This work was supported by the Australian Research Council, Hyper Tech Research Inc OH USA, Alphatech International Ltd, NZ and the University of Wollongong.

**References**

1. T. Takenobu, T. Ito, Dam Hieu Chi, K. Prassides and Y. Iwasa, Phys. Rev. B 64 (2001) 134513.
2. M. Paranthaman, J.F. Thompson, D.K. Christen, Physica C 355 (2001) 1.
3. A. Bharathi, S. Jemina Balaselvi, S. Kalavathi, G.L.N. Reddy, V. Sankara Sastry, Y. Haritharan and T.S. Radhakrishnan, Physica C 370 (2002) 211.
4. J.S. Ahn, E.J. Choi, cond-mat/0103069.
5. I. Maurin, S. Margadonna, K. Prassides, T. Takenobu, T. Ito, D.H. Chi, Y. Iwasa, A. Fitch, Physica B 318 (2002) 392.
6. I. Maurin, S. Margadonna, K. Prassides, T. Takenobu, Y. Iwasa, A.N. Fitch, Chem. Mater. 14 (2002) 3894.
7. W. Mickelson, J. Cumings, W.Q. Han and A. Zettl, Phys. Rev. B 65 (2002) 052505.
8. Zhao-hua Cheng, Bao-gen Shen, Jian Zhang, Shao-ying Zhang, Tong-yun Zhao and Hong-Wu Zhao, J. Appl. Phys. 91 (2002) 7125.
9. R.A. Ribeiro, S. Bud'ko, C. Petrovic, P.C. Canfield, Physica 382 (2002) 166.
10. M. Avdeev, J.D. Jorgensen, Cond-Mat/0301025.
11. S. X. Dou, X. L. Wang, J. Horvat, D. Milliken, A. H. Li, K. Konstantinov, E. W. Collings, M. D. Sumption and H. K. Liu Physica C, 361 (2001) 79.
12. X. L. Wang, S. Soltanian, J. Horvat, A. H. Li, M. J. Qin, H. K. Liu and S. X. Dou, Physica C, 361 (2001) 149.
13. S. X. Dou, S. Soltanian, J. Horvat, X. L. Wang, P. Munroe, S. H. Zhou, M. Ionescu, H. K. Liu and M. Tomsic, Appl. Phys. Lett., 81 (2002) 3419.
14. X.L. Wang, S.H. Zhou, M.J. Qin, P. Munroe, S. Soltanian, H.K. Liu and S.X. Dou, Physica C, 385 (2003) 461.
15. Y. Bugoslavsky, L. F. Cohen, G. K. Perkins, M. Polichetti, T. J. Tate, R. Gwilliam and A. D. Caplin, Nature, 411 (2001) 561.
16. C. B. Eom, M.K. Lee, J. H. Choi, L. J. Belenky, X. Song, L. D. Cooley, M. T. Naus, S. Patnaik, J. Jiang, M. Rikel, A. Polyanskii, A. Gurevich, X. Y. Cai, S. D. Bu, S. E. Babcock, E. E. Helstrom, D. C. Larbalestier, N. Rogado, K. A. Regan, M. A. Hayward, T. He, J. S. Slusky, K. Inumaru, M. K. Hass and R. J. Cava, Nature, 411 (2001) 558.

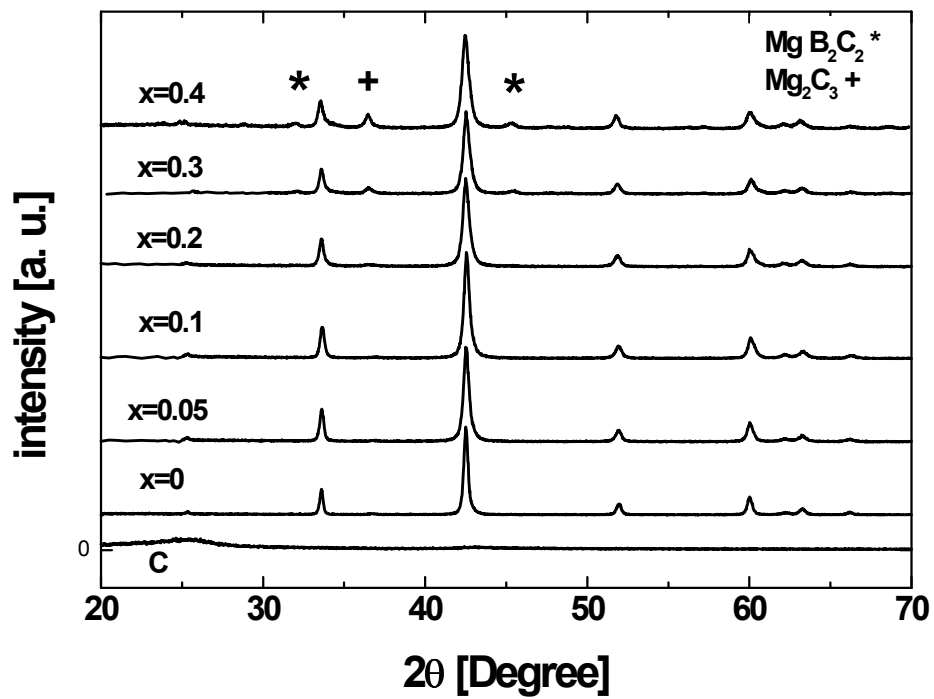

Fig. 1

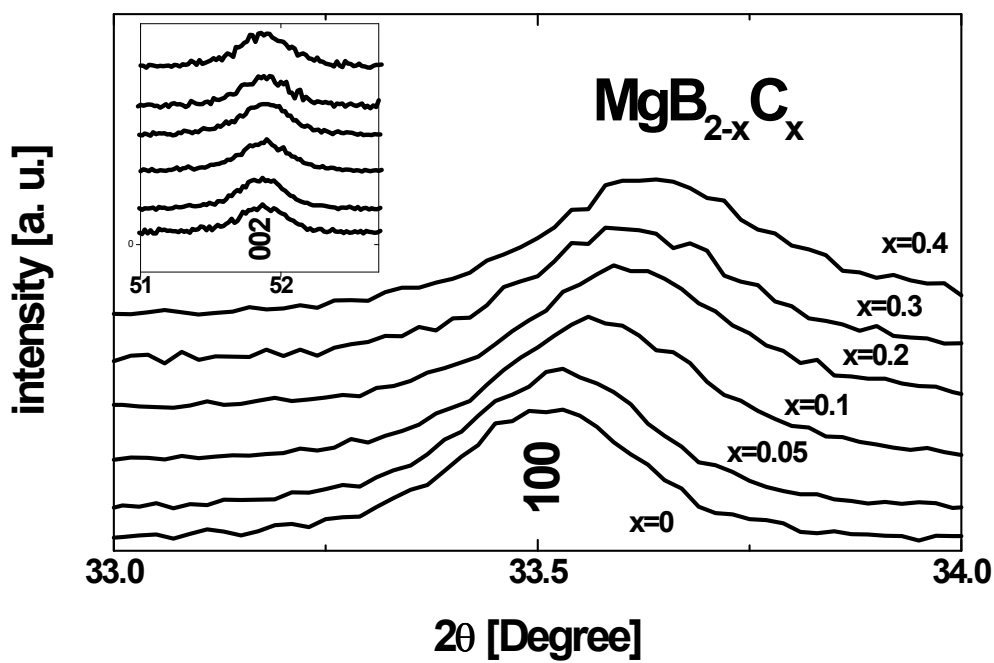

Fig. 2

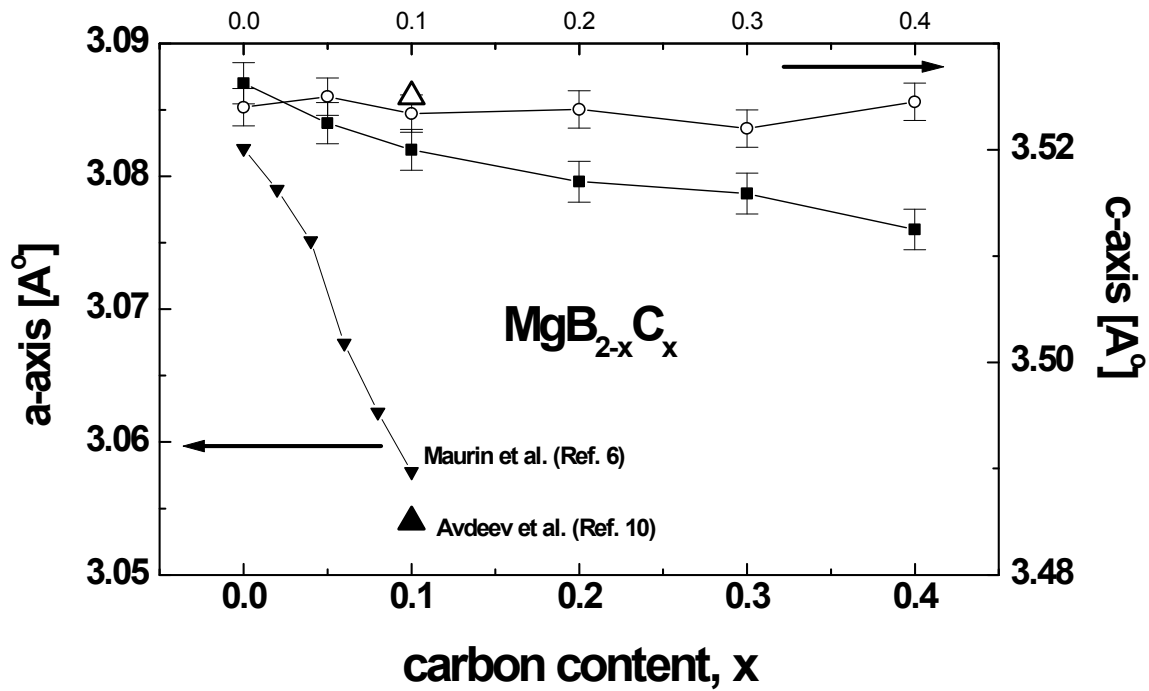

Fig. 3

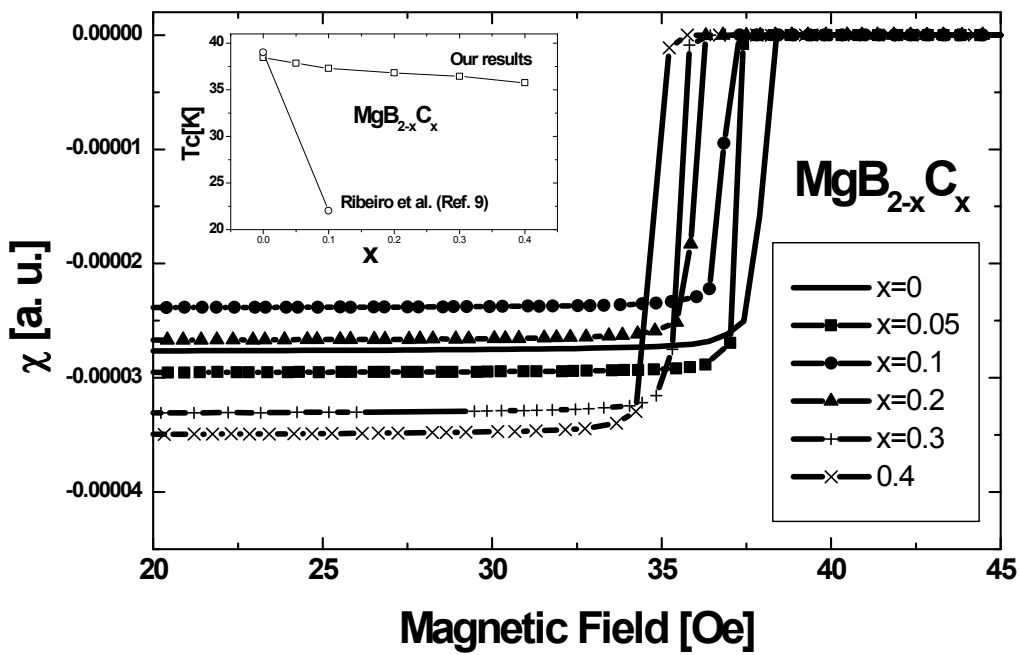

Fig. 4

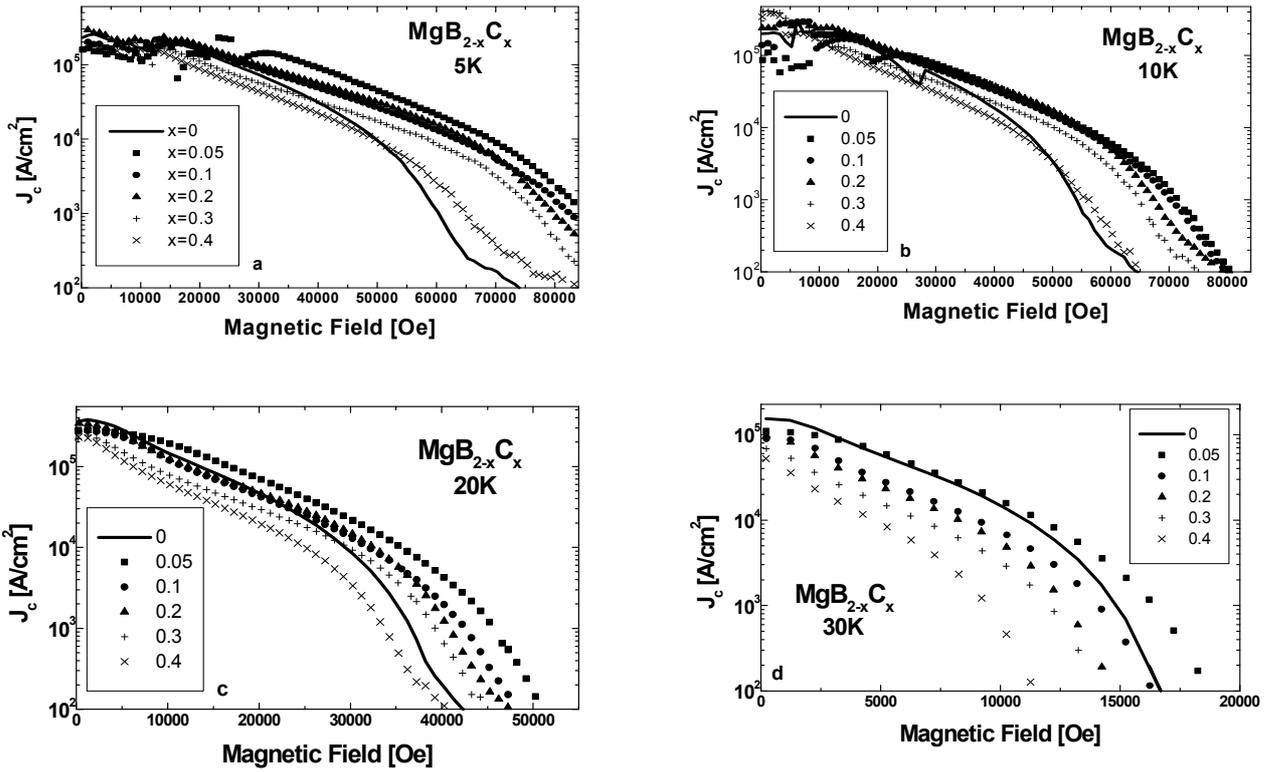

Fig. 5

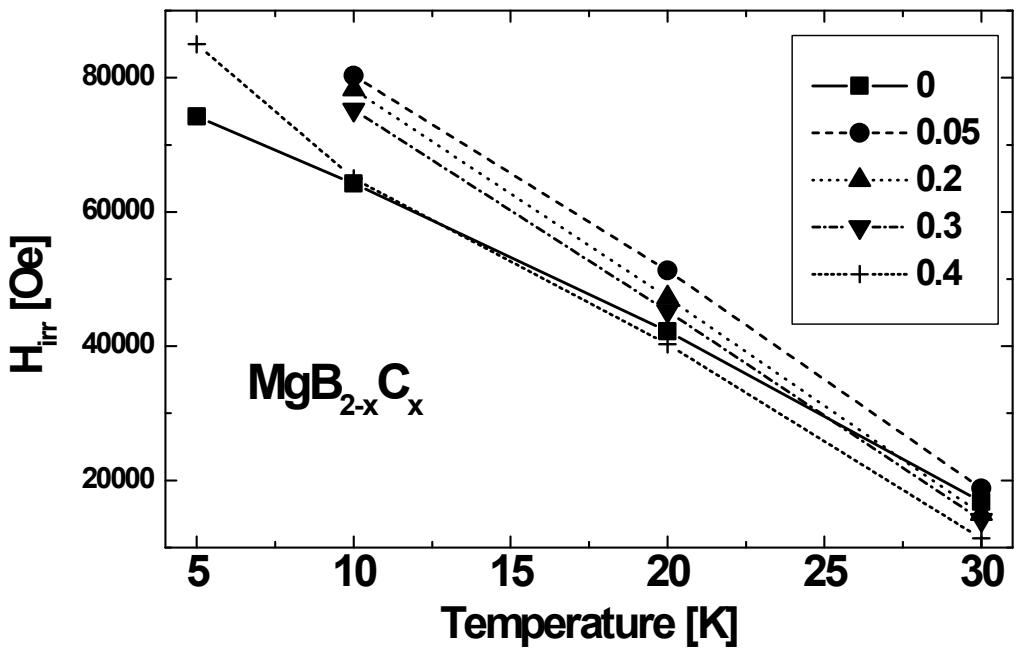

Fig. 6

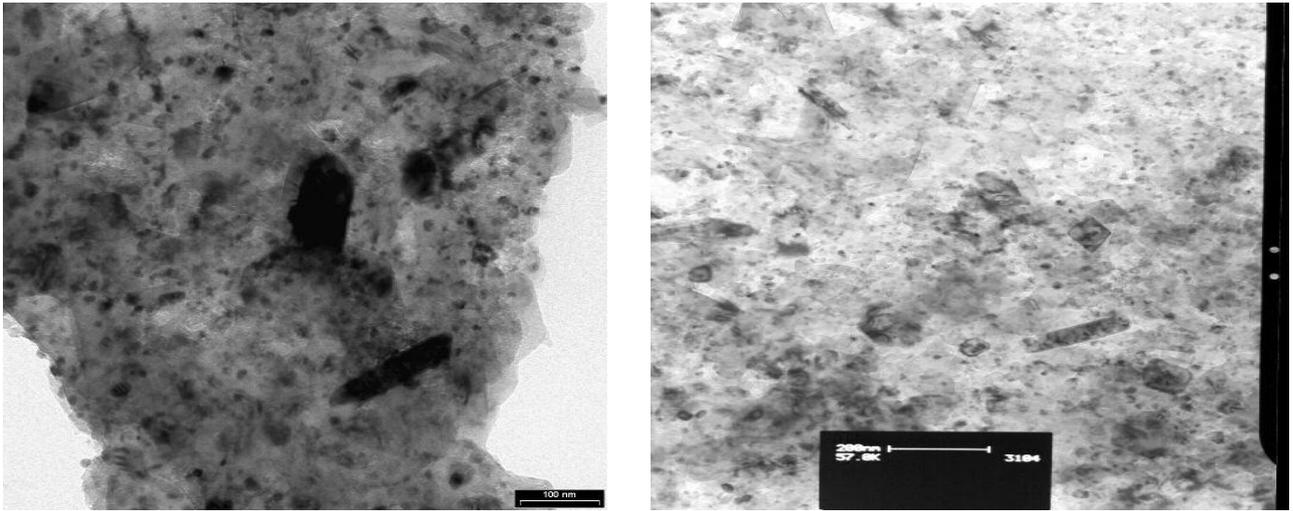

Fig. 7

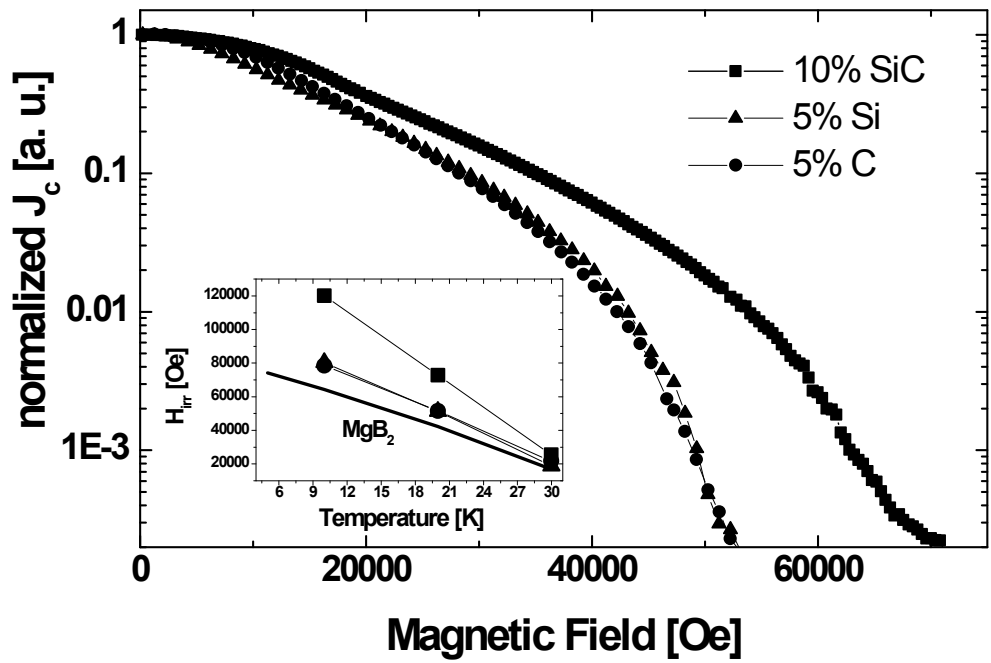

Fig 8

**Figure Captions:**

Fig. 1. XRD patterns of $MgB_{2-x}C_x$ composition for x=0, 0.05, 0.1, 0.2, 0.3 and 0.4 as well as the XRD pattern of the starting C powder.

Fig. 2. The (100) and (002) (inset) Bragg reflections for $MgB_{2-x}C_x$ composition with x= 0, 0.05, 0.1, 0.2, 0.3, and 0.4

Fig. 3. Change in the *a* and *c* lattice parameters in $MgB_{2-x}C_x$ as a function of the nominal C content x. The lattice parameters extracted from the previously published studies by Maurin et al. [6] and Avdeev et al. [10] are also included.

Fig. 4. AC susceptibility (real part) vs. magnetic field for different nominal C content x for $MgB_{2-x}C_x$. The inset shows the $T_c$ changes with x for the same composition including for x=0.1, reported by Ribeiro et al. (Ref. 9).

Fig. 5. The $J_c$ field dependence of $MgB_{2-x}C_x$ composition for x=0, 0.05, 0.1, 0.2, 0.3 and 0.4 at 5K, 10K, 20K and 30K.

Fig. 6 Irreversibility lines for $MgB_{2-x}C_x$ composition for x=0, 0.05, 0.1, 0.2, 0.3 and 0.4.

Fig. 7 TEM images for C doped $MgB_{2-x}C_x$ composition at x=0.05 and 0.1.

Fig. 8. A Comparison of $J_c(H)$ and $H_{irr}$ for SiC, C and Si doped $MgB_2$.